\documentclass{aa55898-25}  

\usepackage{natbib}
\bibpunct{(}{)}{;}{a}{}{,} 

\usepackage{array, multirow, graphicx}
\usepackage[dvipsnames]{xcolor}
\usepackage{wasysym}[integrals]
\usepackage{amssymb}
\usepackage{pifont}
\usepackage{siunitx}
\usepackage{booktabs, tabularx}
\usepackage{cellspace, makecell} 
\usepackage{mathtools}
\usepackage{txfonts}
\usepackage{mathrsfs}
\usepackage{comment}
\usepackage{orcidlink}

\usepackage[]{hyperref}
\definecolor{linkcolor}{rgb}{0.0,0.3,0.5}
 \hypersetup{
     colorlinks=true,
     linkcolor=linkcolor,
     filecolor=linkcolor,
     citecolor = linkcolor,      
     urlcolor=linkcolor,
     }

\newcounter{magicrownumbers}

\begin{document}

\title{Bayesian luminosity function estimation in multi-depth datasets with selection effects: A case study for $3<z<5$ Lyman $\alpha$ emitters}
\titlerunning{Bayesian luminosity function estimation}

\authorrunning{Tornotti et al.}

    \author{Davide Tornotti$\,$\orcidlink{0009-0001-3388-8742}
        \inst{1}\thanks{\href{mailto:d.tornotti@campus.unimib.it}{d.tornotti@campus.unimib.it}},
    Matteo Fossati$\,$\orcidlink{0000-0002-9043-8764} \inst{1,2},
    Michele Fumagalli$\,$\orcidlink{0000-0002-9043-8764} \inst{1,3},  
    Davide Gerosa$\,$\orcidlink{0000-0002-0933-3579} \inst{1,4},
    \\ Lorenzo Pizzuti$\,$\orcidlink{0000-0001-5654-7580} \inst{1}, \and
    Fabrizio Arrigoni Battaia$\,$\orcidlink{0000-0002-4770-6137} \inst{5}
    }

\institute{Dipartimento di Fisica ``G. Occhialini'', Universit\'a degli Studi di Milano-Bicocca, Piazza della Scienza 3, 20126 Milano, Italy \label{unimib}
\and
    INAF – Osservatorio Astronomico di Brera, Via Brera 28, I-21021 Milano, Italy
\and
    INAF – Osservatorio Astronomico di Trieste, Via G. B. Tiepolo 11, I-34143 Trieste, Italy
\and
    INFN, Sezione di Milano-Bicocca, Piazza della Scienza 3, 20126 Milano, Italy
\and 
    Max-Planck-Institut f\"ur Astrophysik, Karl-Schwarzschild-Str. 1,
    D-85748 Garching bei M\"unchen, Germany
   }

\abstract{
We present a hierarchical Bayesian framework designed to infer the luminosity function of any class of object by jointly modelling data from multiple surveys with varying depth, completeness, and sky coverage. Our method explicitly accounts for selection effects and measurement uncertainties (e.g. in luminosity) and can be generalized to any extensive quantity, such as mass. We validated the model using mock catalogues; from this we determined that deep data reaching $\gtrsim 1.5$ dex below a characteristic luminosity ($\tilde{L}^\star$) are essential to reducing biases at the faint end ($\lesssim 0.1$ dex) and that wide-area data help constrain the bright end.
As a proof of concept, we considered a combined sample of 1176 Lyman $\alpha$ emitters at redshift $3 < z < 5$ drawn from several MUSE surveys, ranging from ultra-deep ($\gtrsim 90$~hr) and narrow ($\lesssim 1$ arcmin$^2$) fields to shallow ($\lesssim 5$~hr) and wide ($\gtrsim 20$ arcmin$^2$) fields. With this complete sample, we constrain the luminosity function parameters $\log(\Phi^\star/\mathrm{Mpc^{-3}}) = -2.86^{+0.15}_{-0.17}$, $\log(L^\star/\mathrm{erg\,s^{-1}}) = 42.72^{+0.10}_{-0.09}$, and $\alpha = -1.81^{+0.09}_{-0.09}$, where the uncertainties represent the $90\%$ credible intervals. These values are in agreement with the results of studies based on gravitational lensing that reach $\log(L/\mathrm{erg\,s^{-1}}) \approx 41$, although differences in the faint-end slope underscore how systematic errors are starting to dominate. In contrast, wide-area surveys represent the natural extension needed to constrain the brightest Lyman $\alpha$ emitters [$\log(L/\mathrm{erg\,s^{-1}}) \gtrsim 43$], where statistical uncertainties still dominate.
}

\maketitle
%

\section{Introduction}

The luminosity function (LF) is one of the most fundamental statistical descriptors of the properties of a population of astrophysical objects \citep[e.g.][]{Ciardullo1989, Boyle2000, Miyaji2001, Blanton2003, Kelly2012, Baldry2012, Ajello2012, Bouwens2021}. It is largely used in the context of galaxy formation models, providing a census of the evolution --- or lack thereof --- of galaxy populations across cosmic time. The LF, denoted as $\Phi(L,z)$, is defined as the number of sources per unit comoving volume and luminosity interval. Constraining the parameters that shape the LF and characterizing its evolution with redshift are critical steps in studying the changing demographics of galaxies throughout cosmic history \citep[e.g.][]{Richards2006, Muzzin2013, Herenz2019}.

A fundamental aspect for an accurate determination and parametrization of the LF, in addition to improved observational techniques and higher-quality datasets, lies in the statistical methodology used to infer its parameters. A robust and flexible formalism is even more critical when catalogues are incomplete due to complex selection functions, which themselves depend on the observational techniques adopted. Different methods have been developed to model the LF and can be broadly categorized into non-parametric and parametric approaches (see \citealt{Johnston2011} for a review). 

The former includes several binning techniques, most notably the $1/V_\mathrm{max}$ estimators \citep{Schmidt1968} and the $C^{-}$ method for deriving the cumulative LF \citep{Lynden-Bell1971}. After binning the sources by luminosity, a procedure that carries the arbitrary choices of bin width and phase, a parametric function --- often a Schechter function \citep{Press&Schechter1974, Schechter1976} --- is typically fitted to the binned data using different optimization techniques.
Another parametric approach, known as the maximum likelihood estimator \citep[e.g.][]{Sandage1979}, has the significant advantage of deriving the Schechter function parameters without binning the data, thus preserving the statistical constraining power of the full dataset. However, even this method has limitations and shortcomings: (i) it usually does not directly constrain the LF normalization, but rather only its shape. The normalization is derived by requiring that the expected number of sources matches the number detected in the survey.  (ii) It may not properly account for the selection effects introduced by the observations. (iii) It often does not take into account the uncertainties associated with both the luminosities and redshifts of the sources. Finally, (iv) it generally does not allow for the combination of heterogeneous datasets from different surveys with distinct selection functions and sky coverage. 
Partial attempts to address these limitations include \citet{Kelly2008} for (i), \cite{Marshall1983} and \cite{Drake2017} for (ii), \citet{Chen2003}, \cite{Kelly2008}, and \cite{Mehta2015} for (iii), and \cite{Heyl1997} for (iv). 

In this paper we present a comprehensive analysis strategy that addresses all these limitations at once. Our model is a hierarchical Bayesian framework that we used to estimate the LF parameter; it allows us to combine multiple surveys with an accurate treatment of selection effects and to reconstruct the properties of the parent population, as traced by the observed samples. Specifically, we constructed the likelihood function of the observed data given a set of parameters for an assumed form of the LF, linking the observed quantities to the true, unobservable properties of the sources while explicitly accounting for the survey selection functions. From this, we derived the posterior probability distribution of the LF parameters conditioned on the observed data. We then validated our Bayesian method by generating mock catalogues, both for individual surveys with idealized selection functions and different sky coverage and for combinations of surveys with heterogeneous characteristics. 

Although we focus in this work on the problem of estimating the LF of galaxies, we emphasize that this framework is fully general and can be applied to any population of objects described by an observed luminosity, or even generalized to any extensive quantity, such as mass. Furthermore, the model, here explicitly for a Schechter function, can be generalized to any shape by replacing $\Phi$ with the desired form. 

As a case study, we applied our model to the analysis of high-redshift Lyman $\alpha$ emitters (LAEs), the distribution of which has become a cornerstone in investigations of galaxy formation and evolution, and it traces the large-scale structures in the early Universe \citep[e.g.][]{Ouchi2020, Leclercq2020, Fossati2021, Herrera2025}. The LAE LF has been measured over a wide range of redshifts, from the local Universe up to $z\sim 8$, through various observational programmes since their first detections in the late 1990s. Early studies relied primarily on narrow-band (NB) imaging techniques to identify LAEs. Advances in instrumentation at world-class observatories have enabled NB imaging surveys to become an ideal technique for mapping large sky areas, thus providing initial constraints on the bright end of the LF [$\log(L_{\mathrm{Ly}\alpha}/\mathrm{erg\,s^{-1}})\gtrsim 43$]. However, NB imaging studies are limited in terms of depth and the purity of the LAE selection. 

These limitations can be overcome by spatially resolved spectroscopic surveys mapping the three-dimensional distribution of LAEs in smaller, but representative, volumes. Thanks to the advent of a new class of integral-field spectrographs, such as the Multi Unit Spectroscopic Explorer \citep[MUSE;][]{Bacon2010} on the European Southern Observatory’s Very Large Telescope (VLT), it is now possible to perform sensitive and blind searches for emission-line sources in the range $3\lesssim z \lesssim 6$, enabling a more complete census of spectroscopically confirmed LAEs. During the past decade, several studies with MUSE have constrained the LAE LF down to $\log(L_{\mathrm{Ly}\alpha}/\mathrm{erg\,s^{-1}})\gtrsim 41.4$ \citep[e.g.][]{Drake2017, Herenz2019, Fossati2021}. Using the magnification provided by gravitational lensing by massive foreground galaxy clusters, some efforts have also pushed the detection threshold to the very faint end, reaching $\log(L_{\mathrm{Ly}\alpha}/\mathrm{erg\,s^{-1}})\lesssim 40$ \citep{Thai2023, deLaVieuville2019}. However, these faint-end constraints of the LF are accompanied by increasing uncertainties due to cosmic variance, small-number statistics, and the dependence on complex lensing models. 

In this work we applied our Bayesian model to multiple MUSE surveys, including the two deepest fields currently available: the MUSE Ultra Deep Field (MUDF; \citealt{Lusso2019, Fossati2019}) and the MUSE eXtremely Deep Field (MXDF; \citealt{Bacon2021, Bacon2023}). In addition, we considered the MUSE Analysis of Gas around Galaxies \citep[MAGG;][]{Lofthouse2020, Fossati2021} and the MUSE-Wide \citep[MW;][]{Herenz2019} surveys. From these datasets, we drew a combined sample of 1176 LAEs to constrain the parameters of the Ly$\alpha$ LF over a wide luminosity range ($\approx 2.8$ dex) at $3 < z < 5$.

The structure of this paper is as follows. Section~\ref{sect:bayes-model} introduces our hierarchical Bayesian model with the treatment of multiple surveys. Readers interested only in the general formalism will find all relevant information in this section. To validate the method, we used the explicit example of estimating the Ly$\alpha$ LF of LAEs. In Sect.~\ref{sect:data-red} we describe the MUSE datasets and the construction of the observed LAE catalogues, and in Sect.~\ref{sect:mock-model} we use mock catalogues that simulate real observational conditions to validate the method. Finally, in Sect.~\ref{sect:data-model} we apply our framework to the real combined MUSE datasets, and we compare our results with the literature. We summarize our conclusions in Sect. \ref{sect:conclusions}.

Throughout this paper, we adopt a standard $\Lambda$ cold dark matter cosmology with $\Omega_m = 0.31$ and $H_0 = 67.7 \,\mathrm{km\,s^{-1}\,Mpc^{-1}}$ \citep{Planck2020}. We use $\log(\cdot)$ to indicate base-10 logarithms.

\section{Hierarchical Bayesian framework for multi-depth surveys}\label{sect:bayes-model}

In this section we describe in detail the expanded Bayesian inference formalism we used to model the LF and estimate parameter posteriors, assuming observations characterized by uncertainties on redshift and luminosities and selection biases. This same formalism is discussed by \citet{Loredo2004}, \citet{Mandel2019}, and \citet{Vitale2022}, which we adapted and expanded to our specific case: the study of the LF in multi-depth surveys. 
The formalism described below applies to multi-dimensional data.

In this Bayesian framework, we considered a population of objects characterized by a set of parameters, $\theta,$   that describes their individual properties.
Any population of sources characterized by an extensive quantity could be used, but here we chose to make the formalism explicit for the case of galaxies with luminosity and redshift, such that $\theta = (L, z)$. The distribution of events within our galaxy population is governed by a set of parameters, $\lambda$, such that the number density of objects is given by
\begin{equation}
    \frac{\mathrm{d}N}{\mathrm{d}\theta}(\lambda) = Np_\mathrm{pop}(\theta|\lambda'),
    \label{eq:rate}
\end{equation}
where we have separated the parameters $\lambda$ into the overall normalization factor $N$ (representing the true number of objects in the population, i.e. the number of LAEs in our case) and the subset of parameters $\lambda'$ that describes the shape of the population distribution, $p_\mathrm{pop}(\theta|\lambda').$

For galaxies, the LF can be modelled using a Schechter function, parameterized by $\lambda' = (\tilde{L}^\star, \alpha)$, where $\tilde{L}^\star$ is the characteristic logarithmic luminosity (with typical units of $\mathrm{erg\,s}^{-1}$), and $\alpha$ is the faint-end slope of the LF, 
\begin{equation}
\Phi(\tilde{L} | \lambda', \Phi^\star) = \ln(10)\Phi^\star 10^{(\tilde{L}-\tilde{L^\star})(1+\alpha)}\exp{\left[-10^{(\tilde{L}-\tilde{L}^\star)}\right]}, \label{eq:lum_f}
\end{equation}
which denotes the number of sources per comoving volume and per unit log-luminosity, $\tilde{L}$. The parameter $\Phi^\star$ represents the normalization (usually in $\text{Mpc}^{-3}$), and is related to the total number of LAEs, $N$, which, by definition, is given by the integral of the LF multiplied by the total comoving volume $V_\text{com}^\text{tot}$ (see below).
Here, we considered an LF that is independent of redshift over the considered interval. In its more general form, however, the LF could vary with redshift and adopt a different functional form. 

The distribution of events in the population is sampled by a set of observations, $N_\mathrm{obs}$, with true parameters $\{{\theta_i\}}_{i=1}^{N_\mathrm{obs}}$. This means that, in reality, we do not know the true value of $N$, as our survey contains only $N_\mathrm{obs}$ galaxies. Because of this, we should also consider $N$ as an additional parameter that needs to be estimated. For each observed galaxy in the population, we obtained an uncertain measurement of its parameters, which is characterized by an associated likelihood function $p(\theta^\text{obs}|\theta)$ that relates the observed measurements $\theta_i^\text{obs} = (\tilde{L}_i^\text{obs},z_i^\text{obs})$ to the true event parameters $\theta_i$. We assumed a factorized likelihood made of a log-normal distribution for $L^\text{obs}$ with standard deviation $\sigma_{\tilde{L}}$, and a Dirac $\delta$ function for $z^\text{obs}$. In symbols, one has
\begin{equation}
p(\theta_i^\text{obs} | \theta_i) = \frac{1}{\sqrt{2 \pi }\sigma_{\tilde{L_i}}} \exp \left( - \frac{(\tilde{L}_i - \tilde{L}_i^\text{obs})^2}{2 \sigma_{\tilde{L_i}}^2} \right) \delta(z_i - z_i^\text{obs}).
\end{equation}
The use of a $\delta$ function rests on the assumption of a negligible error in the redshift estimate compared to the redshift interval we considered, which is the case with spectroscopically confirmed catalogues. 

Crucially, we also had to take the selection effects associated with the observations into account. We could only obtain a sample of galaxies that represents a biased subset of the intrinsic population, as some of them are more challenging or even impossible to detect.
Therefore, we had to account for a detection probability $P_\text{det}(\theta, t)$, which depends both on the properties of the objects ($\theta$) and on the limiting sensitivity of the survey (a parameter that is not intrinsic to the population). When focusing on flux-based surveys, as in our examples of the LAE LF, we can use the mean exposure time $t$ as a proxy for the survey depth. This probability can be estimated empirically by analysing the noise properties of the survey and performing source injection and recovery experiments (see an example of this implementation in Sect.~\ref{sect:data-red}).

In the presence of multiple surveys, $N_\text{survey}$, with different mean exposure times $t_k$, we can define a set of selection probabilities $\{P_\text{det,k}(\theta)\}_{k=1}^{N_\text{survey}}$. 
Empirically, for each survey, we can determine a threshold for the luminosity parameter, corresponding to the luminosity at which the selection function is 
complete above a certain threshold (for example, $10\%$), $\tilde{L}_\text{min} = \log(L_\text{min})$. An associated $\Omega_k$, describing the accessible solid angle of the $k$-th survey, must be specified.  
Having done so, we are in a position to define a detection statistic that depends solely on the data, making it possible to split the integration domain into two disjoint sets: one corresponding to the data that yield a detection statistic above a predetermined threshold, and its complement \citep[see e.g.][]{Kelly2008, Taylor2018, Mandel2019, Vitale2022}. 

Our goal is to determine the population properties described by the parameters $\lambda$, but we can only rely on a finite and limited set of observations $\{\theta_i^\text{obs}\}$ subject to selection biases and measurement uncertainties. Therefore, we compute the posterior probability on $\lambda$ using Bayes' theorem 
\begin{equation}
    p(\lambda | \{\theta_i^\text{obs}\}) = \frac{p(\{\theta_i^\text{obs}\}|\lambda)\,\pi(\lambda)}{p(\{\theta_i^\text{obs}\})},
    \label{eq:bayes}
\end{equation}
where $p(\{\theta_i^\text{obs}\}|\lambda)$ is the likelihood of obtaining the observed data given the population parameters, $\pi(\lambda)$ is the prior distribution on $\lambda$ and $p(\{\theta_i^\text{obs}\})$ is the evidence.

We follow the bottom-up derivation by \cite{Mandel2019} and assume that each observation is independent. Furthermore, we assume that the generating process of the observed data is an inhomogeneous Poisson process, meaning that the number of galaxies observed in any given region of parameter space (e.g. luminosity and redshift) is a random realization drawn from a Poisson distribution whose expectation value is determined by the underlying function (e.g. LF) that describes the distribution of objects in the population. This statistical model accounts for the discrete nature of galaxy counts (Poisson) and their variation across luminosity and redshift (inhomogeneous), reflecting the expected physical distribution of galaxies in the Universe. 

In the following, we refer to each individual survey as the $k$-th survey, with $k = 1, \dots, N_\mathrm{survey}$, and to each source detected within that survey as $i_k = 1, \dots, N_{\mathrm{obs},k}$. When considering the combined sample across all surveys, we denote the ensemble of all observed sources by $i = 1, \dots, N_\mathrm{tot}$, where we defined $N_\mathrm{tot} = \sum_{k=1}^{N_\mathrm{survey}} N_{\mathrm{obs},k}$.

We then write the likelihood for the $k$-th survey as 
\begin{equation}
p_k(\{\theta_{i_k}^\text{obs}\}|\lambda', N) \propto \frac{N_\text{det,k}^{N_\text{obs,k}}}{e^{N_\text{det,k}}}\,\prod_{i_k=1}^{N_\mathrm{obs,k}} \frac{\int \mathrm{d}\theta\,p(\theta_{i_k}^\text{obs}|\theta)\,p_\mathrm{pop}(\theta|\lambda')}{\omega_k\int \mathrm{d}\theta P_\mathrm{det,k}(\theta)\,p_\mathrm{pop}(\theta|\lambda')}.
\label{eq:tot_p}
\end{equation}
The expected number of detections $N_\text{det,k}$ is computed as
\begin{equation}
    N_\text{det,k} = \omega_k \int 
    \text{d}\theta P_\text{det,k}(\theta)\,
    N\,p_\mathrm{pop}(\theta|\lambda'),
    \label{eq:Ndet_gen}
\end{equation}
where the factor $\omega_k=\Omega_k/4\pi$ accounts for the limited solid angle of the survey. 
In Eq.~(\ref{eq:tot_p}), by multiplying both the numerator and the denominator inside the product by $N$, we note that the denominator corresponds exactly to $N_\mathrm{det,k}$. Bringing this term outside the product results in $N_\mathrm{det,k}^{N_\mathrm{obs,k}}$, which cancels with the same factor outside the product.

Considering explicitly our case, Eq.~(\ref{eq:rate}) can be expressed as
\begin{equation}
    \frac{\text{d}N}{\text{d}\tilde{L} \, \text{d}z} = N p_{\text{pop}}(\tilde{L}, z | \lambda'),
\end{equation}
where $p_{\text{pop}}(\tilde{L}, z| \lambda')$ represents the normalized distribution of the galaxy population. 
By definition, the LF can be expressed as
\begin{equation}
\Phi(\tilde{L}, V_\text{com} \,|\, \lambda', N) = \frac{\text{d}N}{\text{d}\tilde{L} \, \text{d}V_\text{com}},
\end{equation}
i.e. the number of events in the population we expect in the log-luminosity range $\tilde{L},\tilde{L}+\text{d}\tilde{L}$ and in the comoving volume range $V_\text{com}, V_\text{com}+\text{d}V_\text{com}$.
This yields
\begin{equation}
    p_{\text{pop}}(\tilde{L}, z| \lambda') = \frac{1}{N}\Phi(\tilde{L}, V_\text{com}(z) \,|\, \lambda', N) \frac{\text{d}V_\text{com}}{\text{d}z}(z),
    \label{eq:Ppop}
\end{equation}
where $N$ is, by definition, the total number of sources in the population.
In the equations above,
\begin{equation}
    \frac{\text{d}V_\text{com}}{\text{d}z}(z) = \frac{4\pi c}{H_0}\frac{D_L(z)^2}{(1+z)^2E(z)}
    \label{eq:comV}
\end{equation}
is the differential comoving volume, where $E(z)$ contains the cosmology terms, $D_L(z)$ is the luminosity distance, $c$ is the speed of light, $H_0$ is the Hubble constant (see \citealt{Hogg2000}), and we have accounted for the total solid angle $4\pi$. 
If we assume a simplified version of the LF without explicit dependence on $z$, as in Eq.~(\ref{eq:lum_f}), we can write
$N = \Phi^\star\,I(\tilde{L}^\star, \alpha)$, where we have explicitly factorized $\Phi^\star$ as a normalization constant (without any $z$ dependence) out of the integral over $\tilde{L}$ in Eq.~(\ref{eq:lum_f}), with
\begin{equation}
    I(\tilde{L}^\star, \alpha) = V_\text{com}^\text{tot} \int \text{d}\tilde{L}\,\ln(10)\,10^{(\tilde{L}-\tilde{L^\star})(1+\alpha)}\exp{\left[-10^{(\tilde{L}-\tilde{L}^\star)}\right]}. 
\end{equation}
With the simplification we operated on Eq. (\ref{eq:tot_p}), the likelihood can be written as
\begin{align}
    p_k(\{\tilde{L}_{i_k}^\text{obs}\}, \{z_{i_k}^\text{obs}\}|\lambda', N) &\propto e^{-N_\text{det,k}} 
    \prod_{i_k=1}^{N_\mathrm{obs,k}} \int \int
    \text{d}\tilde{L} \, \text{d}z \, \notag \\
    & \times p(\tilde{L}_{i_k}^\text{obs}, z_{i_k}^\text{obs}|\tilde{L}, z) \,
    N \, p_\mathrm{pop}(\tilde{L}, z|\lambda')
\end{align}
with
\begin{equation}
    N_\text{det,k} = \frac{\Omega_k}{4\pi} N{\int \int \text{d}\tilde{L}\text{d}z\, P_\mathrm{det,k}(\tilde{L}, z)\,p_\mathrm{pop}(\tilde{L}, z|\lambda')}\:.
    \label{eq:Ndet}
\end{equation}

Finally, we used the definition in the Eq. (\ref{eq:Ppop}) and the equivalence between the normalization parameter $N$ and $\Phi^\star$ described above to express the likelihood in terms of $\Phi^\star$ instead of $N$:
\begin{align}
p_k(\{\tilde{L}_{i_k}^\text{obs}\}, \{z_{i_k}^\text{obs}\}|\lambda', \Phi^\star) \propto  
e^{-N_\text{det,k}} \prod_{i_k=1}^{N_\mathrm{obs,k}} 
&\int \text{d}\tilde{L}\,
\mathcal{N}(\tilde{L}_{i_k}^\text{obs}|\tilde{L}; \sigma_{\tilde{L}_{i_k}} ) \notag \\
&\hspace{-2cm} \times \Phi(\tilde{L} \,|\, \lambda', \Phi^\star)\frac{\text{d}V_\text{com}}{\text{d}z}(z_{i_k}^\text{obs}),
\label{eq:likelihood}
\end{align}
where we have also accounted for the Dirac delta in $p(\tilde{L}_{i_k}^\text{obs},z_{i_k}^\text{obs}|\tilde{L}, z)$ and used
\begin{equation}
 \mathcal{N}(\tilde{L}_{i_k}^\text{obs}|\tilde{L}; \sigma_{\tilde{L}_{i_k}} ) = \frac{1}{\sqrt{2 \pi }\sigma_{\tilde{L}_{i_k}}} \exp \left( - \frac{(\tilde{L}_{i_k} - \tilde{L}_{i_k}^\text{obs})^2}{2 \sigma_{\tilde{L}_{i_k}}^2} \right).
\end{equation}
The expected number of detected sources in Eq. (\ref{eq:Ndet}) can be expressed as
\begin{equation}
    N_\text{det,k} = \int \text{d}\tilde{L}\Phi(\tilde{L}|\lambda', \Phi^\star)V_\text{com,k}^\text{eff}(\tilde{L}).
\end{equation}
Furthermore, $V_\text{com,k}^\text{eff}(\tilde{L})$ accounts for selection effects and is expressed as
\begin{equation}
   V_\text{com,k}^\text{eff}(\tilde{L}) = \frac{\Omega_k}{4\pi} \int \text{d}z\frac{\text{d}V_\text{com}}{\text{d}z}(z)P_\mathrm{det,k}(\tilde{L}, z).
   \label{eq:vmax}
\end{equation}
It is crucial to note how the selection effects are introduced only in the denominator factor of Eq.~(\ref{eq:tot_p}) \citep[see][]{Loredo2004} and in the Poissonian term in Eq.~(\ref{eq:likelihood}).
Since each survey can be considered independent of the other (they are disjoint in area), the final likelihood can be expressed as the product
\begin{equation}
 p(\{\theta_i^\text{obs}\}|\lambda', \Phi^\star) \propto \prod_{k=1}^{N_\text{survey}} p_k(\{\theta_{i_k}^\text{obs}\}|\lambda', \Phi^\star).
\end{equation}

We implemented this formalism by using the \texttt{UltraNest} package \citep{Buchner2021}, which is a Python implementation of nested sampling \citep{Skilling2004}. This algorithm efficiently produces posterior samples along with the evidence.
As input, the model requires the different catalogues of galaxies from the various surveys, along with the associated selection functions and the area of the sky covered. In our implementation, for the luminosity threshold below which sources are not considered, we adopted the limit at which the selection function reaches $10\%$ completeness, similar to other works in the literature \citep[e.g.][]{deLaVieuville2019, Fossati2021}. 
To assess the impact of the chosen threshold, we also applied the model adopting $5\%$ and $20\%$ completeness limits (see Sect.~\ref{sect:data-model}).

Having laid out the expanded Bayesian formalism that describes the LF in the presence of multi-depth surveys with uncertainties, we proceeded by validating the method and applying it to a concrete example: the study of the LF for LAEs. In the following section we introduce the main LAE datasets we incorporated in this work. Readers not interested in, or already familiar with, these technical aspects can resume reading from Sect.~\ref{sect:mock-model}.

\section{Observations and sample selection}\label{sect:data-red}

The LAEs analysed in this paper originate from four different surveys: (i) the MUDF \citep{Lusso2019, Fossati2019}, (ii) the MXDF \citep{Bacon2021, Bacon2023}, (iii) the MAGG \citep{Lofthouse2020, Fossati2021}, and (iv) the MW survey \citep{Herenz2019}. The LAE catalogue and the determination of the selection function for the MAGG and MW surveys are presented by \cite{Fossati2021} and \cite{Herenz2019}, respectively, and those of the MW survey are publicly available. In this work, we provide a detailed discussion of the construction of the LAE catalogue and the determination of the selection functions for the MUDF survey, and briefly describe the properties of the other samples in the following subsections. 

For each dataset, we considered LAEs within the redshift range $3 \leq z \leq 5$, where the MUSE sensitivity is highest. Applying this selection criterion and the individual luminosity thresholds at which the completeness is $10\%$, we obtained the final catalogues, which comprise a total of $1176$ sources. Their properties are summarized in Table~\ref{tab:samples_prop}.

\begin{table}[ht]
    \centering
    \caption{Main properties of the surveys used in this work.}
    \label{tab:samples_prop}
    \begin{tabular}{lcccc}
        \toprule
        Survey & $t_\mathrm{exp}$  & $Area$ & $\log(L_\mathrm{min})$ & $N_\mathrm{obs}$\\
        & (hr) & (arcmin$^2$) & $(\mathrm{erg\,s^{-1}}$) & \\
        \midrule
        MXDF & $91-140$ & $0.9$ & 40.89 & 139\\
        MUDF 1 & $91-120$ & $0.5$ & 41.03 & 48 \\
        MUDF 2 & $48-91$ & $0.5$ & 41.11 & 40 \\
        MUDF 3 & $19-48$ & $0.5$ & 41.29 & 38 \\
        MUDF 4 & $10-19$ & $0.5$ & 41.49 & 23\\
        MUDF 5 & $3-10$ & $1.2$ & $41.63$ & 21 \\
        MAGG  & $\approx 5$ & $26.5$ & 41.69 & 686 \\
        MW  & $\approx 1$ & $22.2$ & 42.13 & 181 \\
        \bottomrule
    \end{tabular}
    \tablefoot{$N_\mathrm{obs}$ is the number of sources above the reported luminosity threshold $\log(L_\mathrm{min})$.}
\end{table}

\subsection{The MUDF dataset}
The MUDF is a 142-hour VLT/MUSE programme covering an area of approximately $2.1 \times 1.9$ arcmin$^2$ in a field that hosts two bright quasars at $z \approx 3.22$, reaching the highest sensitivity ($\gtrsim 90$ hr) in the $\approx 0.55$ arcmin$^2$ central region. Details of observations and data reduction are provided by \citep{Lusso2019, Fossati2019, Tornotti2025a}. In this work, we utilized the full-depth dataset presented by \citet{Tornotti2025a}. Specifically, we adopted the final datacube in which the quasar and stars point spread functions and the extended continuum sources have been subtracted using a non-parametric algorithm \citep[e.g.][]{Borisova2016, ArrigoniBattaia2019}.

The procedure used to detect compact LAEs follows a similar approach to that described in, for example, \citet{Lofthouse2020} and  \citet{Fossati2021}, and is briefly summarized here.
First, we ran the Spectral Highlighting and Identification of Emission (SHINE) code \citep{Fossati2025}, following an approach similar to that by \citet{Tornotti2025b} for the extraction of Ly$\alpha$ extended filamentary emission, but adopting appropriate parameters for compact sources: (i) a signal-to-noise ratio (S/N) threshold per voxel of $3$ with spatial Gaussian smoothing kernel of $\sigma=2$ pixels; (ii) a minimum number of $27$ connected voxels; (iii) a minimum of $3$ spectral channels (i.e. $>3.75\,\text{\AA}$) along at least one spatial pixel; and (iv) a group of connected voxels spanning no more than $50$ wavelength channels ($62.5~\AA)$, to reject residuals from continuum sources.  

The candidate line emitters are then classified into two credible levels ($1$ and $2$) based on the integrated signal-to-noise ratio ($ISN$), corrected for the uncertainty covariance \citep{Lofthouse2020}. Specifically, class 1 includes sources with $ISN > 7$, ensuring high purity but lower completeness, while class 2 includes sources with $5 < ISN < 7$, improving completeness at the expense of purity. To construct a high-purity sample, we consider only class 1 sources hereafter. All candidates are visually inspected by three authors (DT, MF, and MFo) independently. During this process, we also examined the shape of the segmentation map and the extracted spectrum of each emitter to identify possible skyline residuals or other artefacts.

Following this procedure, a total of $212$ LAEs have been extracted and confirmed. For each confirmed emitter, the redshift is determined from the peak of the Ly$\alpha$ emission line in the spectrum. In cases where a double-peaked profile is present and no non-resonant line is detected, the red peak is used to assign the redshift.
We then estimated the total flux of the Ly$\alpha$ line. LAEs can be decomposed into a bright compact core and a faint, diffuse halo \citep[e.g.][]{Wisotzki2016, Leclercq2017}. As a result, the flux obtained directly from the segmentation map typically underestimates the total flux, as it misses the faint extended emission in the outskirts of the halos. To this end, we adopted a curve-of-growth (CoG) approach using circular apertures centred on each source \citep{Fossati2021}. We applied this technique to pseudo-NB images, obtained by collapsing the datacube within $\pm 15\,\text{\AA}$ around the source redshift. The total flux of each object is then defined from the analysis of the cumulative CoG profile. 
This procedure is performed interactively, by inspecting CoG diagnostic plots to ensure that the measured flux is indeed dominated by the LAE emission. In a few cases, where necessary, we masked artefacts, residuals, or the edges of the field by limiting the apertures or the pseudo-NB accordingly.
As a final step, we computed the luminosity from the fluxes using the luminosity distance at the redshift of the LAE within our cosmological model, after correcting for Milky Way dust extinction using the calibrated extinction map by \citet{Schlafly2011} and applying the extinction law by \citet{Fitzpatrick1999}.

To study the LF of LAEs, we needed to model the selection function of the MUDF, i.e. the probability of detecting a LAE with a given luminosity and redshift in our datacube. To this end, we injected mock sources into the datacube and tested their detectability using the same procedure described above.
To better represent the morphological properties of LAEs, we injected observed LAEs \citep[see][]{Herenz2019, Fossati2021}. Specifically, we selected $11$ LAEs with $ISN>25$ in the MUDF dataset, spanning the redshift range $3 < z < 5$ that is representative of the entire LAE population. For each selected source we extracted a sub-cube of $25$x$25$ spatial pixels and $17$ spectral pixels ($21.25~\AA$) and normalized the data to the total flux. We used these bright sources as non-parametric and nearly noise-free LAE templates. They were then randomly selected and injected in the real datacube after being rescaled to a random flux value. This approach is the best method for simulating observed LAEs across the full range of fluxes that can be observed in MUSE datacubes.
Due to the high $ISN$, the impact of the residual noise in the sub-cubes used as LAE models is negligible, even when rescaled to high flux levels.

Next, we computed the recovery fraction of the injected sources in bins of redshift and flux. Converting these fluxes into luminosities for each redshift bin yields the final selection function, which is modulated primarily by the MUSE sensitivity and the impact of bright sky emission lines. Since the MUDF exposure map decreases quasi-radially from the central deep region (up to $120$~hr) towards the edges ($\lesssim 10$~hr), we incorporated this spatial dependence into the selection function. During the injection process, we tracked the exposure time at the position of each mock source. In the subsequent analysis, we computed distinct selection functions corresponding to each depth-defined region. 

We divided the MUDF field into five regions characterized by exposure times of $91-120$, $48-91$, $19-48$, $10-19,$ and $3-10$ hours. These bins were chosen to maximize the LAE statistics, reflecting the depth variation within the datacube.
In the following, these five regions (named MUDF 1, MUDF 2, MUDF 3, MUDF 4, and MUDF 5) are treated as independent surveys, each characterized by its own selection function. The luminosity at which the selection function reaches 10\% completeness is, for each region, $\log(L_{\mathrm{min}}/\mathrm{erg\,s^{-1}}) = 41.03$, $41.11$, $41.29$, $41.49$ and $41.63$.

\subsection{The MXDF dataset}
The catalogue of line emitters from the MXDF is publicly available and presented in \cite{Bacon2023}; hence, we describe only the salient properties here. The MXDF is a 155-hour VLT/MUSE programme covering an area of approximately $2.45~\mathrm{arcmin}^2$, located within the Hubble Ultra Deep Field \citep[HUDF;][]{Beckwith2006}, and reaching high sensitivity ($\gtrsim 90$ and up to $140$~hours) in the central $\approx 0.9~\mathrm{arcmin}^2$ region. Details of observations and data reduction are provided by \citet{Bacon2023}. These authors have also released an emission line catalogue of all sources in the datacube, from which we selected the LAEs.

To ensure the consistency of the analysis methods in this work, we 
extracted the candidate LAEs from the continuum-subtracted MXDF cube using the same procedure as done for the MUDF, including the estimate of each candidate's $ISN$. We then cross-matched these LAEs with the available MXDF catalogue to obtain a final list of individual sources with $ISN>7$. All sources meeting this criterion are already included in the original sample in \citet{Bacon2023}.  
For each source, we estimated the total flux using the same CoG method. We then computed the associated selection function by running similar injection and recovery simulations. Finally, we considered only the ultra-deep region ($\gtrsim 90$ hr) to derive the final selection function. This area covers nearly twice the ultra-deep region of the MUDF, allowing us to build a sample of LAEs that further constrains the faint end of the LF. The luminosity at which the selection function reaches a completeness of $10\%$ is $\log(L_{\mathrm{min}}/\mathrm{erg\,s^{-1}}) = 40.89$.

\subsection{The MAGG dataset}

The MAGG survey is a VLT/MUSE programme targeting 28 quasar fields at $z \approx 3.2$-$4.5$, each observed for a total of approximately 4 hours. Details of the observations and data reduction are provided by \citet{Lofthouse2020, Fossati2021}, along with a description of the extraction and creation of the LAE catalogue and the associated selection function for realistic LAEs. In summary, the strategy adopted for this dataset is consistent with the one described for the MUDF. Since the exposure time of each field is nearly identical, the survey can be treated as a single large-area programme covering $\approx 26.5~\mathrm{arcmin}^2$, with a selection function reaching 10\% completeness at $\log(L_{\mathrm{min}}/\mathrm{erg\,s^{-1}}) = 41.69$.

\subsection{The MW dataset}
The MW survey is a VLT/MUSE programme consisting of 24 adjacent $1 \times 1~\mathrm{arcmin}^2$ pointings in the CANDELS/Deep region of the GOODS-South field. The total survey area is $22.2~\mathrm{arcmin}^2$, accounting for a $4''$ overlap between individual pointings. Details of the observations and data reduction are provided by \citet{Herenz2019}, along with the extraction and construction of the LAE catalogue and the associated selection function.
Although the extraction process adopts a slightly different methodology, the provided selection function refers to this final catalogue and therefore allows us to include this dataset consistently in our analysis. Furthermore, the luminosity of each source is carefully estimated and corrected to be consistent with a manual CoG analysis adopted in the other datasets, as described in \citet{Herenz2017}. The luminosity at which the selection function reaches 10\% completeness is $ \log(L_{\mathrm{min}}/\mathrm{erg\,s^{-1}}) = 42.13$.

\section{Testing the Bayesian formalism with mock catalogues}\label{sect:mock-model}

In this section we numerically test the Bayesian model using mock catalogues generated from a given LF, demonstrating that we can recover the input parameters from the synthetic sample. To concretely demonstrate the flexibility of this formalism in working on multi-depth datasets, we performed this test not in general terms, but by specializing to the case of an LF for LAEs in surveys that share the characteristics introduced in the previous section. The outcome of this test is easily generalized to any galaxy survey. 
To work on this specific case, we implemented the formalism described above, further assuming (in Eq.~\ref{eq:bayes}) a uniform prior $\pi$ in the parameters $\Phi^\star$, $\tilde{L}^\star$ and $\alpha$ in the ranges $[10^{-4}, 10^{-2}], [41.5, 43.5]$ and $[-2.5, -1]$. As output, we obtain the posterior samples for $\Phi^\star, \tilde{L}^\star$, and $\alpha$, the three parameters that define the Schechter function (see Eq.~\ref{eq:lum_f}). 

For our test, we constructed three mock catalogues that represent different observational scenarios, considering a selection function that depends only on luminosity, thereby neglecting, at this stage, the wavelength (or redshift) dependence. These three catalogues are defined as follows: 
\begin{itemize} 
    \item A shallow survey ($t_{\rm obs} \sim5$ hr with MUSE) covering a wide area of $\sim 25\,\text{arcmin}^{2}$ (MAGG-like), with a selection function characterized by a $10\%$ detection threshold at $\log(L_\mathrm{min}/\mathrm{erg\,s^{-1}})\approx 41.7$; 
    \item A deep survey ($t_{\rm obs} \sim 100$ hr with MUSE) over an area of $\sim 0.5~\text{arcmin}^{2}$ (MUDF 1-like), with a selection function characterized by a $10\%$ detection threshold at $\log(L_\mathrm{min}/\mathrm{erg\,s^{-1}})\approx 41$; 
    \item A combination of the two mock surveys described above. 
    \end{itemize}
We also performed an initial test on a complete mock survey with an area $30$ times larger than that of the full MUDF, which allowed us to constrain both the faint and bright ends of the LF, in order to verify that the Bayesian model can correctly recover the true parameters. 

We assumed that the mock population follows the Schechter LF in Eq.~(\ref{eq:lum_f}), with the parameters inferred for LAEs by \cite{Herenz2019}:
$\log(\Phi^\star/\mathrm{Mpc^{-3}}) = -2.71,
\,\, \log(L^\star/\mathrm{erg\,s^{-1}}) = 42.66, \,\, \alpha = -1.84.$
This choice mimics an LF that is close to a real observed case, but the results of this test are not dependent on the specific parameter values chosen.
We sampled the corresponding probability distribution function of the population [Eq.~(\ref{eq:Ppop})], generating sources with associated redshifts in the range $3 < z < 5$ and luminosities satisfying $\log(L/\mathrm{erg\,s^{-1}}) \geq 40.5.$
While we did not introduce measurement errors in the redshift values, we included a relative uncertainty of $10\%$ in the luminosities.
We then applied a selection function modelled as a sigmoid,
\begin{equation}
S(\tilde{L}) = \frac{1}{1 + \exp\left(-\frac{\tilde{L} - \tilde{L}_{\text{inf}}}{\delta_{\tilde{L}}} \right)},
\end{equation}
defined by the inflection point $\tilde{L}_{\text{inf}}$ that depends on the mock survey and a fixed slope parameter $\delta_{\tilde{L}} = 0.1$, which simulates the sharp decline in sensitivity observed in the real data. We chose an $\tilde{L}_{\text{inf}}$ value that reproduces the $10\%$ detection thresholds described above: $\tilde{L}_{\text{inf}} = 41.86$  ($\tilde{L}_{\text{inf}} = 41.25$) for the shallow (deep) survey. Following this procedure, we generated a sample of $1082$ ($105$) LAEs with luminosities $\log(L/\mathrm{erg\,s^{-1}}) \gtrsim 41.7$ ($\log(L/\mathrm{erg\,s^{-1}}) \gtrsim 41$) for the shallow (deep) survey.

\begin{figure}
\centering
\includegraphics[scale=0.55]{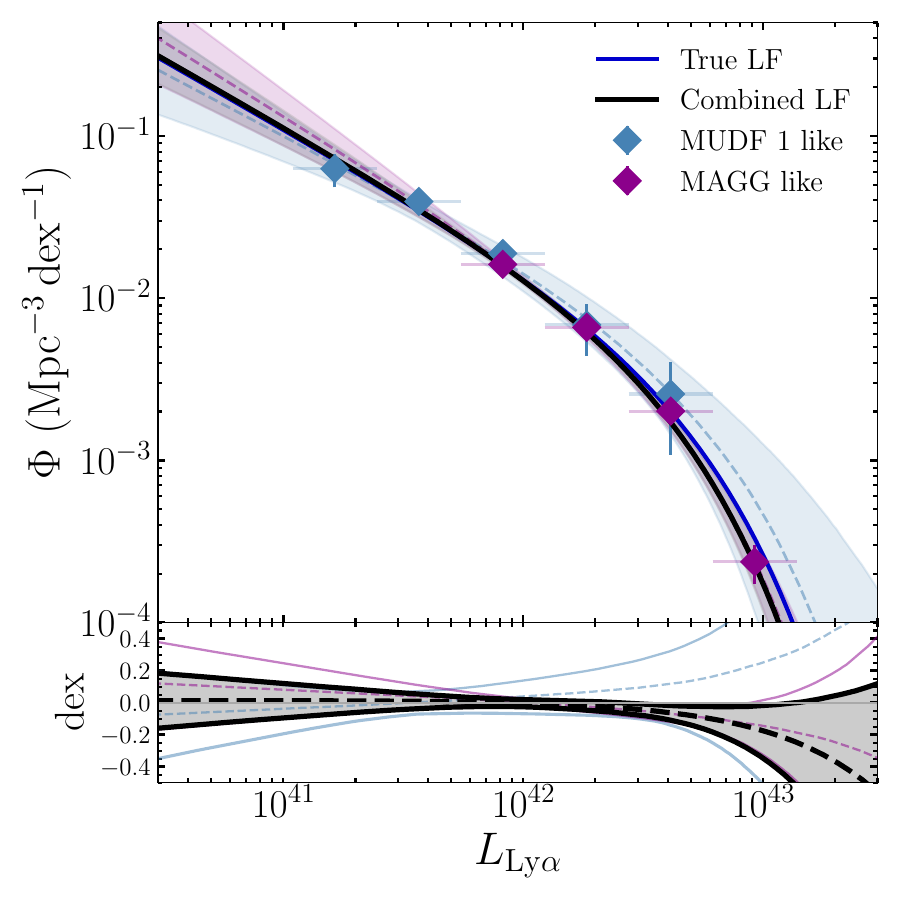}
\caption{Results of the Bayesian model applied to mock catalogues for both individual surveys and their combination. This test demonstrates how combining multiple surveys with varying depths helps constrain the LF with reduced statistical uncertainty across a wider dynamical range of luminosities. In the \textit{top panel} the solid dark blue line indicates the true LF used to generate the mock catalogues. The light blue (magenta) diamonds represent the $1/V_\mathrm{max}$ estimator of the LF obtained from the MUDF 1- (MAGG)-like survey, along with the median of the posterior samples from the Bayesian model (dashed line) and the corresponding $90\%$ credible interval. The solid black line shows the median when combining the two surveys, with the associated $90\%$ credible interval. Horizontal bars on the diamonds represent the bin widths, while vertical bars are the Poisson errors associated with the $1/V_\mathrm{max}$ estimator. In the \textit{bottom panel} the dashed lines (same colours as in the top panel) show the logarithmic difference between the median LF obtained from the Bayesian model and the true LF. The solid lines indicate the $5$th and $95$th percentiles relative to the true LF.}
\label{fig:LF_mock}
\end{figure}

We applied the formalism to these mock samples and compare in Fig.~\ref{fig:LF_mock} the true input LF with the median posterior estimates derived from our Bayesian model for each mock survey scenario. In this figure, we also show the non-parametric $1/V_\mathrm{max}$ LF estimator originally proposed by \citet{Schmidt1968} and later modified to account for redshift- and luminosity-dependent selection functions \citep{Fan2001, Herenz2019}. Within this formalism, and using the same notation as above, the differential LF for the $k$-th survey can be approximated in luminosity bins according to the relation
\begin{equation}
    \Phi(\langle \tilde{L} \rangle)_k = \frac{1}{\Delta \tilde{L}} \sum_{i=1}^{N_\mathrm{obs,k}^{\mathrm{bin}}} \frac{1}{V^\mathrm{eff}_\mathrm{com,k}(\tilde{L}_i)},
\end{equation}
where $V^\mathrm{eff}_\mathrm{com,k}(\tilde{L}_i)$ is the effective comoving volume accessible to the $i$-th galaxy in the $k$-th survey and in the luminosity bin, accounting for the survey's selection function (see Eq.~\ref{eq:vmax}). Here, $\Delta \tilde{L}$ is the width of the luminosity bin, which is set to be $0.35$ dex in Fig. \ref{fig:LF_mock}, while $\langle \tilde{L} \rangle$ is the average Ly$\alpha$ luminosity within that bin. The associated uncertainty in each bin can be estimated following, for example, \citet{Johnston2011} as
\begin{equation}
    \sigma(\langle \tilde{L} \rangle)_k = \sqrt{ \frac{1}{\Delta \tilde{L}^2} \sum_{i=1}^{N_\mathrm{obs,k}^{\mathrm{bin}}} \frac{1}{V^\mathrm{eff}_\mathrm{com,k}(\tilde{L}_i)^2} }.
\end{equation}

The first conclusion we draw from Fig.~\ref{fig:LF_mock} is that the methodology laid out above faithfully recovers the parameters of the input LF. Indeed, by combining a wide and a deep survey to explore a considerable luminosity range with appropriate statistics, we observe that the recovered LF reproduces the input one. However, inspection of the recovered LF from the combined surveys shows a small discrepancy at the bright end [$\log(L/\mathrm{erg\,s^{-1}})\approx 43$]. This is due to residual statistical uncertainty caused by the limited number of luminous LAEs.  

It is also instructive to explore the impact that specific surveys have on the recovery of the LF by considering two typical cases that are frequently encountered in observations: a deep, narrow-area survey and a shallow, wide-area survey. 
We adopted luminosity uncertainties that are consistent with the typical values and the luminosity dependence observed in data.
Specifically, we assumed a fractional uncertainty that decreases with increasing luminosity and ranges from $25\%$ for low-luminosity sources to $2\%$ for bright ones. We also estimated the potential error introduced when extrapolating the model beyond the dynamical range covered by the data.
We first considered a survey with an area of $\approx 26$~arcmin$^2$ and a $10\%$ completeness luminosity limit of $\log(L_\mathrm{min}/\mathrm{erg\,s^{-1}})\approx 41.5$, i.e. approximately $1.2$ dex below the input $\tilde{L}^\star$. The method constrains the true LF at the faint end within an uncertainty of $\approx 0.25$ dex, down to a luminosity of $\log(L/\mathrm{erg\,s^{-1}})\approx 41$, and with a much smaller uncertainty of $\approx 0.02$ dex up to $\log(L/\mathrm{erg\,s^{-1}})\approx 42.9$. Despite the lower statistical uncertainties resulting from the relatively high number of luminous objects in the survey area, the recovered parameters show discrepancies at the level of $\gtrsim 0.1-0.2$~dex, due to the lack of sufficient low luminosity galaxies and the increasing luminosity uncertainties affecting those that are detected, which reduce their statistical weight and favour a steeper faint-end slope.

The opposite effect is noticeable when considering an ultra-deep, small-area survey covering $\approx 0.5$ arcmin$^2$ and reaching a $10\%$ completeness luminosity limit of $\log(L_\mathrm{min}/\mathrm{erg\,s^{-1}})\approx 41$. In this case, the faint end is recovered with high accuracy, $\lesssim 0.01$ dex, down to $\log(L/\mathrm{erg\,s^{-1}})\lesssim 41$. However, the ability to constrain the bright end decreases, reaching $0.25$ dex at $\log(L/\mathrm{erg\,s^{-1}})\approx 43$. 
This finding highlights that the sheer number of objects at specific low luminosities, rather than the quality of individual datasets, drives the accuracy with which we recover the LF faint-end parameters. Indeed, at the cost of a slightly larger discrepancy on $\tilde{L}^\star$ ($\approx 0.1$ dex), the number density set by $\Phi^\star$ and the faint-end slope ($\alpha$) are still well constrained. Higher statistical errors persist due to the limited number of bright sources in such a small-area survey.

We conclude that, in order to reduce uncertainties in the determination of the LF, it is more effective to carry out a deeper survey reaching a $10\%$ completeness luminosity limit $\gtrsim 1.5$ dex below the true $\tilde{L}^\star$ over a small area — enough to obtain only tens of galaxies around $\tilde{L}^\star$. However, this strategy does not allow us to constrain the bright end of the LF (at $\log(L/\mathrm{erg\,s^{-1}})\gtrsim 43$) accurately. A wide-area ($\gtrsim 30-35$ arcmin$^2$) and shallow survey is required as supplemental input. Our formalism natively allows the combination of various surveys, thus enabling the combination of various datasets effortlessly.  

In summary, we have shown that the framework we developed reliably recovers the true LF parameters and provides reliable determinations of the LF in a variety of configurations and survey combinations, fully leveraging the available information.

\section{Application to the LF of LAEs at $3<z<5$} \label{sect:data-model}

\begin{figure}
\centering
\includegraphics[scale=0.44]{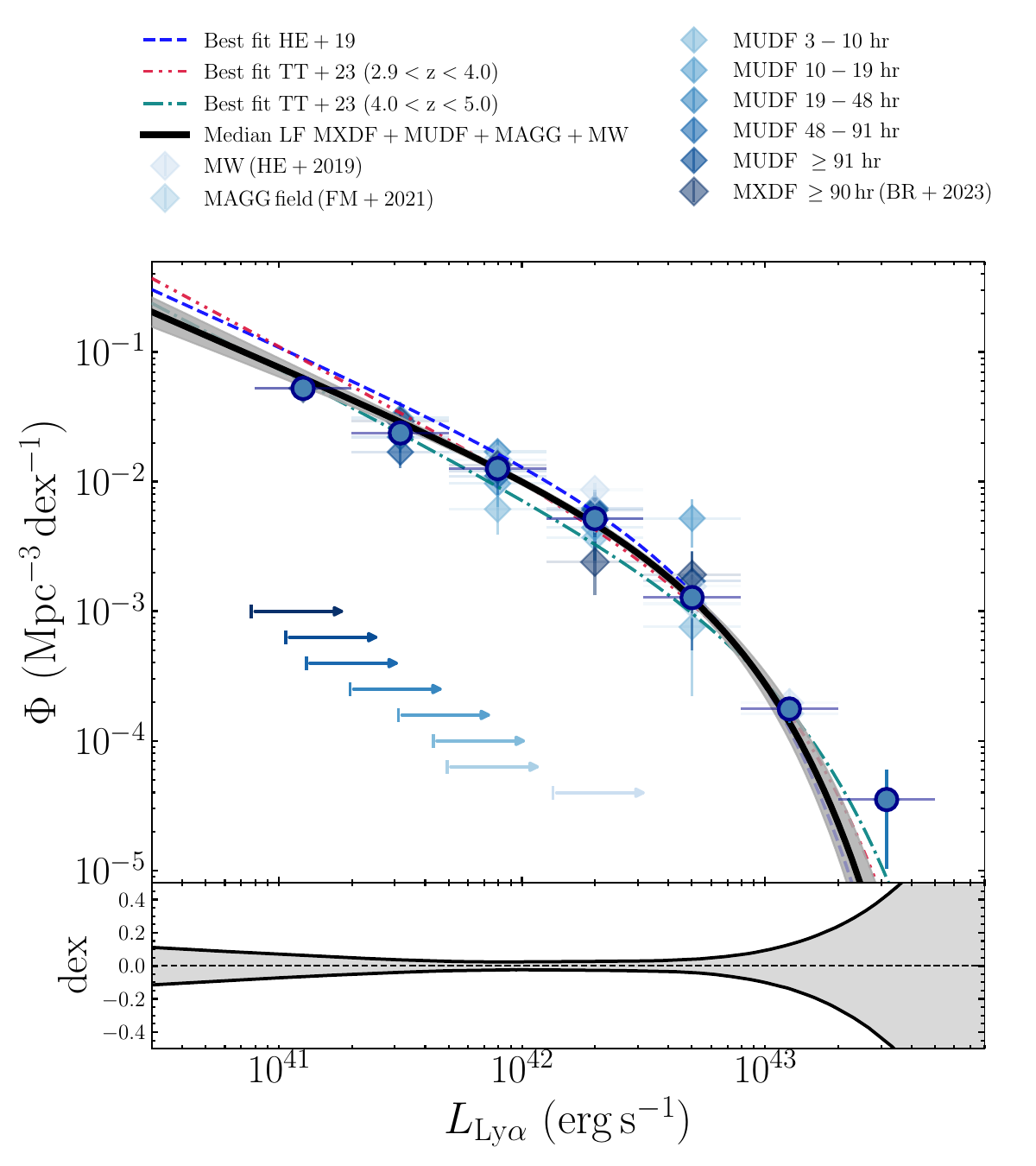}
\caption{LF obtained from the combination of the four MUSE surveys considered in this study (see Table~\ref{tab:samples_prop}). \textit{Top panel}: Median LF reconstructed using our Bayesian model applied to the full sample (solid black line), along with the corresponding $90\%$ credible interval. The blue colour-coded diamonds represent the $1/V_\mathrm{max}$ estimates from each individual survey, with vertical bars indicating the Poisson uncertainties and horizontal bars the bin widths. The dark blue circles show the weighted average of the $1/V_\mathrm{max}$ points across surveys in each bin. Arrows in the same colour scheme indicate the luminosity limit above which each survey reaches a completeness higher than $10\%$. For comparison, coloured lines with different line styles display the best-fit Schechter LFs from various studies in the literature \citep{Herenz2019, Thai2023}. \textit{Bottom panel:} Statistical uncertainty on the LF as a function of luminosity. The solid black lines show the relative $5$th and $95$th percentiles of the Bayesian posterior distribution.}
\label{fig:LF_MAGG+MUDF+MW+MXDF}
\end{figure}

\renewcommand{\arraystretch}{1.4}
\begin{table}[t]
\centering
\caption{Schechter function parameters in different redshift bins.} 
\begin{tabular}{lccc}
\toprule
Redshift & $\log(\Phi^\star/\mathrm{Mpc^{-3}})$ & $\log(L^\star/\mathrm{erg\,s^{-1}})$ & $\alpha$ \\
\midrule
$3 < z < 5$ & $-2.86^{+0.15}_{-0.17}$ & $42.72^{+0.10}_{-0.09}$ & $-1.81^{+0.09}_{-0.09}$ \\
$3 < z < 4$ & $-2.74^{+0.17}_{-0.20}$ & $42.68^{+0.13}_{-0.11}$ & $-1.68^{+0.11}_{-0.11}$ \\
$4 < z < 5$ & $-3.11^{+0.30}_{-0.39}$ & $42.81^{+0.22}_{-0.16}$ & $-2.00^{+0.15}_{-0.13}$ \\
\bottomrule
\end{tabular}
\label{tab:LF_params}
\tablefoot{The values correspond to the medians of the one-dimensional marginalized posterior distributions, and the uncertainties represent the 5th and 95th percentiles.}
\end{table}

\renewcommand{\arraystretch}{1.4}
\begin{table}[t]
\centering
\caption{Schechter function parameters when varying the completeness threshold, $C_\mathrm{th}$.}
\begin{tabular}{lcccc}
\toprule
$C_\mathrm{th}$ & $N_\mathrm{obs}$ & $\log(\Phi^\star/\mathrm{Mpc^{-3}})$ & $\log(L^\star/\mathrm{erg\,s^{-1}})$ & $\alpha$ \\
\midrule
$5\%$ & 1197 & $-2.89^{+0.15}_{-0.17}$ & $42.73^{+0.10}_{-0.09}$ & $-1.83^{+0.09}_{-0.08}$ \\
$10\%$ & 1176 & $-2.86^{+0.15}_{-0.17}$ & $42.72^{+0.10}_{-0.09}$ & $-1.81^{+0.09}_{-0.09}$ \\
$20\%$ & 1133 & $-2.87^{+0.15}_{-0.17}$ & $42.72^{+0.10}_{-0.09}$ & $-1.80^{+0.09}_{-0.09}$ \\
\bottomrule
\end{tabular}
\label{tab:LF_params_ths}
\tablefoot{The values correspond to the medians of the one-dimensional marginalized posterior distributions for the full sample in the redshift range $3<z<5$, and the uncertainties represent the 5th and 95th percentiles.}
\end{table}

Having validated the method with mock data, we applied our Bayesian framework to the MUSE surveys described in Sect.~\ref{sect:data-red}. In parallel, for comparison, we also computed the non-parametric LF using binned estimates with a bin width of $\Delta \tilde{L} = 0.4$ dex for each survey. The binning procedure accounts for the varying luminosity limits of the different surveys, ensuring that no bin includes contributions from surveys whose detection threshold lies within the bin range. This prevents a (possibly severe) underestimation of the number density in the bin that would otherwise require a very large and highly uncertain statistical correction.

The LF obtained from applying the Bayesian model to the combined dataset (MXDF+MUDF+MAGG+MW) is shown in the top panel of Fig.~\ref{fig:LF_MAGG+MUDF+MW+MXDF}, together with the non-parametric $1/V_\mathrm{max}$ estimates from individual surveys and their average, weighted by the number of objects per bin. 
These $1/V_\mathrm{max}$ values and their weighted mean lie in proximity to the inferred parametric model and serve as visual guidance for the model performance.
Arrows indicate, for each survey, the luminosity limit above which the completeness exceeds $10\%$. This helps visualize the minimum luminosity covered by each dataset, highlighting their constraining power in the combined model. 

Below $\log(L/\mathrm{erg\,s^{-1}}) \lesssim 40.9$, the inferred LF is no longer directly constrained by the data. Conversely, at the high-luminosity end, we cannot define a single cutoff point in the constraining power of the adopted datasets. However, due to the volumes probed by the largest surveys we used (MAGG and MW), the model is no longer constrained by the data at $\log(L/\mathrm{erg\,s^{-1}}) \gtrsim 43.4$. 
The bottom panel shows the statistical uncertainty on the LF as a function of luminosity, expressed as the $5$th and $95$th percentiles of the posterior relative to the best-fit. Uncertainties vary from $\lesssim 0.01$ dex at $\log(L/\mathrm{erg\,s^{-1}}) \approx 42$, to $\approx 0.1$ dex at $\log(L/\mathrm{erg\,s^{-1}}) \approx 41$, and increase to $\gtrsim 0.15$ dex for $\log(L/\mathrm{erg\,s^{-1}}) \gtrsim 43$, reflecting the lower number of bright sources. These values are consistent with the findings of the mock tests presented above. The apparent offset between the $1/V_\mathrm{max}$ point in highest-luminosity bin and the Bayesian fit arises from the very low number statistics in that bin (only two sources), which consequently has a limited impact on constraining the LF in the Bayesian model, as also indicated by the larger statistical uncertainty. Additional large volume datasets, can provide stronger constraints on the bright-end of the LF, as further discussed in Sect. \ref{sect:discuss-literature}.

The median values and credible intervals of the Schechter function parameters obtained for the overall sample are reported in Table \ref{tab:LF_params}. We also assessed the dependence of the results on the adopted completeness threshold, considering the luminosities at which the completeness exceeds $5\%$ and $20\%$. 
We applied the model to the redshift range $3<z<5$ and report the associated number of sources above the new luminosity threshold, along with the median values and credible intervals, in Table~\ref{tab:LF_params_ths}. The sharp decline in the completeness results in a variation in the luminosity limits of $\approx 0.04$ dex (on average). This does not significantly impact the inferred LF parameters that remain in good agreement with those estimated at a limit $10\%$.  We believe that the initial choice of $10\%$ provides a good balance between extended minimum luminosity and a sufficient number of observed galaxies without the need of
excessive completeness corrections. 

While we assumed a perfectly known $P_\mathrm{det,k}(\tilde{L},z)$, we assessed the effect of uncertainties on this quantity to the recovery of the LF by generating a set of perturbed realizations of the selection functions shown in Fig. \ref{fig:sel_funcs}. Specifically, we perturbed the transition region of the sigmoid-like completeness curve along the luminosity axis at fixed redshift, introducing Gaussian variations from a few percent up to several tens of percent. This approach allowed us to test the sensitivity of the Bayesian model in the regime with the steepest variations in $P_\mathrm{det,k}(\tilde{L},z)$ with luminosity and, thus, the one potentially more affected by systematics. We notice that even the strongest perturbations produce only marginal changes in the faint-end slope, within the statistical uncertainties. Our inference is therefore statistically robust against reasonable biases and random uncertainties in the selection functions.

\begin{figure}
\centering
\includegraphics[scale=0.45]{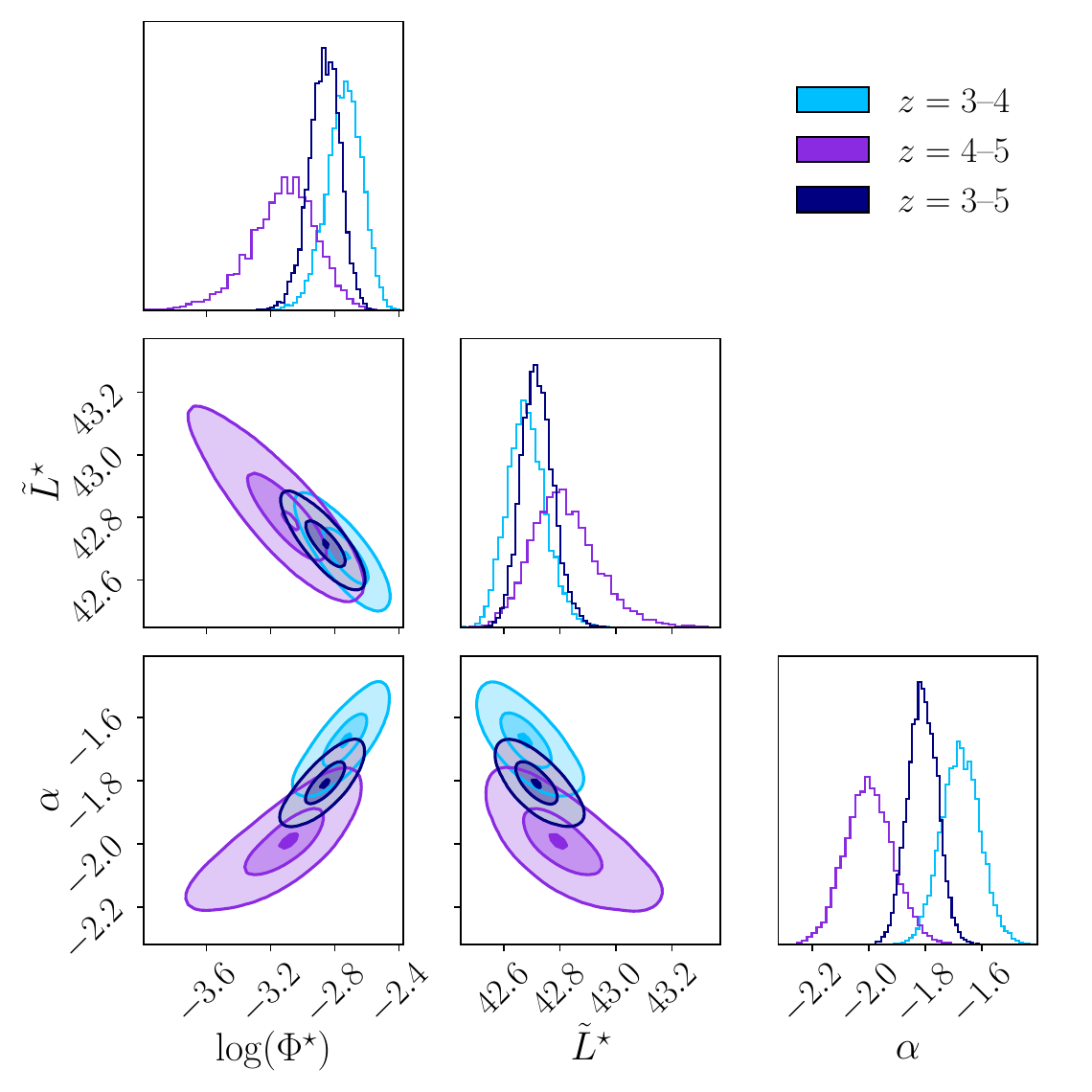}
\caption{Posterior distributions of the LF parameters $\alpha$, $\tilde{L}^\star$, and $\log(\Phi^\star)$ inferred from the Bayesian model for the overall sample (dark blue) and for two redshift bins: $3 < z < 4$ (light blue) and $4 < z < 5$ (purple). \textit{Rightmost panels}: Marginalized one-dimensional posterior distributions for each parameter. \textit{Other panels}: Joint two-dimensional posterior distributions, highlighting the correlations between parameters. The $5$th, $50$th, and $95$th percentiles are shown.}
\label{fig:Redshift_bins}
\end{figure}

Finally, we split the sample into two redshift bins ($3 < z < 4$ and $4 < z < 5$) to test whether the data imply a redshift evolution of the LF. In Fig.~\ref{fig:Redshift_bins}, we show the posterior distributions of the parameters inferred from the Bayesian model in the two redshift intervals, along with the full redshift range we explored. The median marginalized values are reported in Table \ref{tab:LF_params}. We observe a trend in $\Phi^\star$, consistent with a decreasing number density of sources towards a higher redshift. There is a weak but not statistically significant increase in $\tilde{L}^\star$ with redshift, in line with the results of \citet{Thai2023} and \citet{Drake2017}. Instead, the faint-end slope ($\alpha$) shows a stronger evolution, ranging from $-2.00^{+0.15}_{-0.13}$ in the $4 < z < 5$ bin to $-1.68^{+0.11}_{-0.11}$ in the $3 < z < 4$ bin. At higher redshift, we must note that the credible contours show a degeneracy between $\alpha$ and $\tilde{L}^\star$. This arises as a consequence of the smaller number of available objects that can pin down the knee of the LF, allowing for an ambiguity between a shallower LF with a fainter knee or a steeper LF with a brighter $\tilde{L}^\star$.

\subsection{Discussion and comparison with the literature} \label{sect:discuss-literature}

We compared our median LF reconstructed from the posterior samples with those from previous studies, some of which are shown as dashed lines in Fig.~\ref{fig:LF_MAGG+MUDF+MW+MXDF}. In particular, we considered the LF estimates of \citet{Thai2023}, who leveraged strong gravitational lensing by 17 galaxy clusters to probe the faint end of the LF ($\log(L/\mathrm{erg\,s^{-1}}) \lesssim 41$) in the redshift bins $2.9 < z < 4$ and $4 < z < 5$. Their results show overall agreement with our median LF, although with a steeper faint-end slope ($\alpha \approx -2.00$ compared to our median $\alpha = -1.81^{+0.09}_{-0.09}$, considering the overall redshift range $3 < z < 5$).
When comparing the same redshift intervals, we find that our faint-end slope is in agreement within the uncertainties in the $4 < z < 5$ range, but is lower by approximately 15\% in the $3 < z < 4$ bin. Conversely, in the $4 < z < 5$ interval, our slope is steeper by about 19\%, while in the $3 < z < 4$ range it remains higher, but still consistent within $1\sigma$ with the values reported by \citet{deLaVieuville2019}, a  study also exploiting the lensing effect to reach low luminosities ($\log(L/\mathrm{erg\,s^{-1}}) \lesssim 41$).

\citet{Thai2023} include all sources with a completeness above 1\%, thereby incorporating a larger number of objects that may be more strongly affected by completeness corrections. This is also particularly relevant because magnification models become increasingly uncertain at the lowest luminosities, where sources lie close to the lensing caustic lines and are thus highly sensitive to their precise positions relative to these lines (see \citet{Thai2023} for a detailed discussion). In our analysis, we adopted a more conservative completeness threshold of 10\%, which was explicitly incorporated into the Bayesian model used to determine the median LF. Indeed, the total number of expected sources (or the volume density) is treated as a free parameter and is modelled accordingly, taking this threshold into account.
Moreover, \citet{Thai2023} reports that applying a 10\% completeness cut results in a shallower faint-end slope—particularly in the $3 < z < 4$ bin, where they find a value of approximately $\alpha \approx -1.78$, which is closer to our estimate. A steeper slope can indeed be expected when a small number of objects carry a very large statistical correction weight. 
The generally lower values of the faint-end slope reported by \citet{deLaVieuville2019} who applied a $10\%$ completeness threshold can instead be attributed to larger statistical uncertainties due to the smaller sample size, as also discussed in \citet{Thai2023}.

We also show in Fig. \ref{fig:LF_MAGG+MUDF+MW+MXDF} the best-fit obtained in the work of \citet{Herenz2019}, who used the MW data included in our combined fit. Although the luminosity range covered by these data is limited to $\log(L/\mathrm{erg\,s^{-1}}) \gtrsim 42.1$, and is therefore subject to larger statistical uncertainties, their best-fit is in good agreement with our median LF once combined with the deeper datasets, although we observe a slightly lower $\tilde{L}^\star$ and higher $\Phi^\star$. 

We conclude that, because of the ability offered by our Bayesian framework to combine multiple surveys with varying observational characteristics, we are now in a position to robustly constrain the shape of the LF of $3 < z < 5$ LAEs across three dex in luminosity. In particular, the availability of sufficient numbers of deep surveys has now considerably reduced the statistical uncertainties on the faint-end slope. While this is welcome news, comparison with literature values highlights how we are moving towards a regime in which systematic effects -- such as those imposed by completeness corrections -- start to be relevant. Further improvements in the determination of $\alpha$ should invest in high-quality datasets where purity and completeness are well characterized, rather than increasing sample sizes or pushing to even fainter luminosities\footnote{Ultra-deep experiments retain values, including the mitigation of cosmic variance from pencil beam surveys and the testing of uncharted regions of parameter space, such as deviations from a single power law or even the existence of a cutoff in the LF.}. In contrast, there is an ample margin for improving the statistical significance with which the bright end of the LF is constrained. At high luminosities, the LAE surveys used in this work lack the statistical power to probe the LF where the expected number density is $\Phi < 10^{-4}~\rm Mpc^{-3}~dex^{-1}$. Much larger volumes can be harvested from the MUSE archive, an effort that is currently only limited by the speed with which LAE catalogues can be assembled. Alternatively, wide-area surveys from other instruments \citep[e.g. HETDEX;][]{Hill2008} can offer interesting datasets that complement the MUSE surveys. With our work, the tool for stitching together various datasets is now available. 

\section{Summary and conclusions} \label{sect:conclusions}
We have presented a hierarchical Bayesian model to constrain the shape of the LF, accounting for the selection effects of different surveys, such as completeness, luminosity limits, and limited sky coverage. By combining heterogeneous data to fully exploit the constraining power of the wide dynamical range spanned by all diverse available surveys, this approach provides a robust way to determine the parameters of the underlying population of sources, once a functional form (e.g. a Schechter function) is assumed. In addition, the model incorporates measurement uncertainties on the physical properties associated with each source (e.g. luminosity).

While this model is very general and can be applied to describe the distribution of any extensive quantity for a class of astrophysical object, in this work we focused on the case study of establishing the LF of galaxies. In particular, we applied our framework to derive the LF of LAEs using data from four large MUSE surveys, ranging from ultra-deep observations ($\gtrsim 90$ hr) over small areas ($\lesssim 1$ arcmin$^2$) to shallower exposures ($\lesssim 5$ hr) over much larger fields ($\gtrsim 20$ arcmin$^2$). When combined, these surveys account for nearly $500$~hr of MUSE observing time. Before applying the model to real data, we validated its performance using mock catalogues, assessing its ability to recover the input parameters and investigating how statistical uncertainties on the inferred parameters depend on the properties of the input surveys.

We summarize our main conclusions as follows: 
\begin{itemize}
    \item The hierarchical Bayesian model developed in this work incorporates several novel key points for the estimate of the LF: It enables the combination of multi-depth surveys, each characterized by its own selection effects \citep[see e.g.][]{Mandel2019}. It explicitly treats the normalization factor as a free parameter of the underlying population, accounting for it through a Poisson term in the likelihood, and includes measurement uncertainties on the observed properties (e.g. the galaxies' luminosity).
    
    \item Tests of the model using mock catalogues of LAEs suggest that it is highly beneficial to extend the observations to luminosities $\gtrsim 1.5$ dex below the characteristic luminosity ($\tilde{L}^\star$) 
    even over relatively small areas. 
    These deep observations can be effectively complemented by wide-area but shallower surveys, which help constrain the bright end of the LF.

    \item Through a population of 1176 LAEs obtained from the combination of the ultra-deep MXDF and MUDF surveys and the large-area MAGG and MW surveys, we studied the LF both in the redshift range $3<z<5$ and in the redshifts bins $3<z<4$ and $4<z<5$. This allowed us to constrain the faint-end slope of the LF with statistical errors $\lesssim 0.1$ dex at $\log(L/\mathrm{erg\,s^{-1}}) \approx 41$. Assuming the population of LAEs is described by a Schechter function, the median values along with the $90\%$ credible interval obtained from the one-dimensional marginalized posterior in the redshift range $3<z<5$ are: $\log(\Phi^\star) = -2.86^{+0.15}_{-0.17}$,  $\log(L^\star/\mathrm{erg\,s^{-1}}) = 42.72^{+0.10}_{-0.09}$, and $\alpha = -1.81^{+0.09}_{-0.09}$.

    \item The median LF is consistent with previous results of past studies that constrained the faint end \citep[e.g.][]{deLaVieuville2019, Thai2023}, although our median value of $\alpha$ is $\approx 10 \%$ lower than those obtained by \citet{Thai2023} and $\approx 6 \%$ higher than \citet[see the text for a more detailed discussion]{deLaVieuville2019}. The steepening of the faint-end slope with redshift, suggested also by \citet{deLaVieuville2019} and \citet{Thai2023}, is tentatively confirmed with this sample.

\end{itemize}

We show that, within our Bayesian framework and by leveraging a large and ultra-deep MUSE sample, it is in principle possible to recover the true LAE LF within $\lesssim 0.05 - 0.1$ dex, with statistical errors $\lesssim 0.1$ dex up to $\log(L/\mathrm{erg\,s^{-1}}) \gtrsim 43$. This is particularly true at the faint end, where we have demonstrated that the model applied to these types of datasets can, in principle, constrain the underlying LF, provided that the sample selection process is not affected by significant biases.

While in this work we focused on MUSE surveys, there are other  large-area observational efforts in the literature (e.g. the LAE sample from the HETDEX survey; \citealt{Hill2008, Zhang2021}) that can further constrain the bright end of the LF [$\log(L/\mathrm{erg\,s^{-1}}) \gtrsim 43$], a regime where our current sample still yields estimates that are affected by  larger statistical uncertainties, or where deviations from a single Schechter function are possible \citep{Spinoso2020}. 
In the future, upcoming and proposed instruments like BlueMUSE \citep[][]{Richard2019} and WST \citep[][]{Mainieri2024} will expand both the accessible redshift range (down to $z\approx2$) and the depth and number density of the samples, providing tighter constraints on the shape and evolution of the LAE LF. 
The ability to fully capture uncertainties and selection effects in the model is therefore essential to characterizing the galaxy population when combining data from different observational facilities, and represents a crucial step towards building an accurate picture of galaxy evolution across cosmic time.

\begin{acknowledgements}
This work is supported by the Italian Ministry for Research and University (MUR) under Grant 'Progetto Dipartimenti di Eccellenza 2023-2027' (BiCoQ). 
D.G. is supported by  ERC Starting Grant No.~945155--GWmining, 
Cariplo Foundation Grant No.~2021-0555, 
MSCA Fellowship No.~101064542--StochRewind,
MSCA Fellowship No.~101149270--ProtoBH,
MUR PRIN Grant No.~2022-Z9X4XS, 
MUR Young Researchers Grant No. SOE2024-0000125
and the ICSC National Research Centre funded by NextGenerationEU. 
\end{acknowledgements}


\bibliographystyle{aa55898-25}
\bibliography{aa55898-25} 

\newpage 
\begin{appendix}  

\raggedbottom

\section{Selection functions of the surveys}

\begin{figure}[h!]
\centering
\includegraphics[scale=0.42,trim={0 2.5cm 0 4cm},clip]{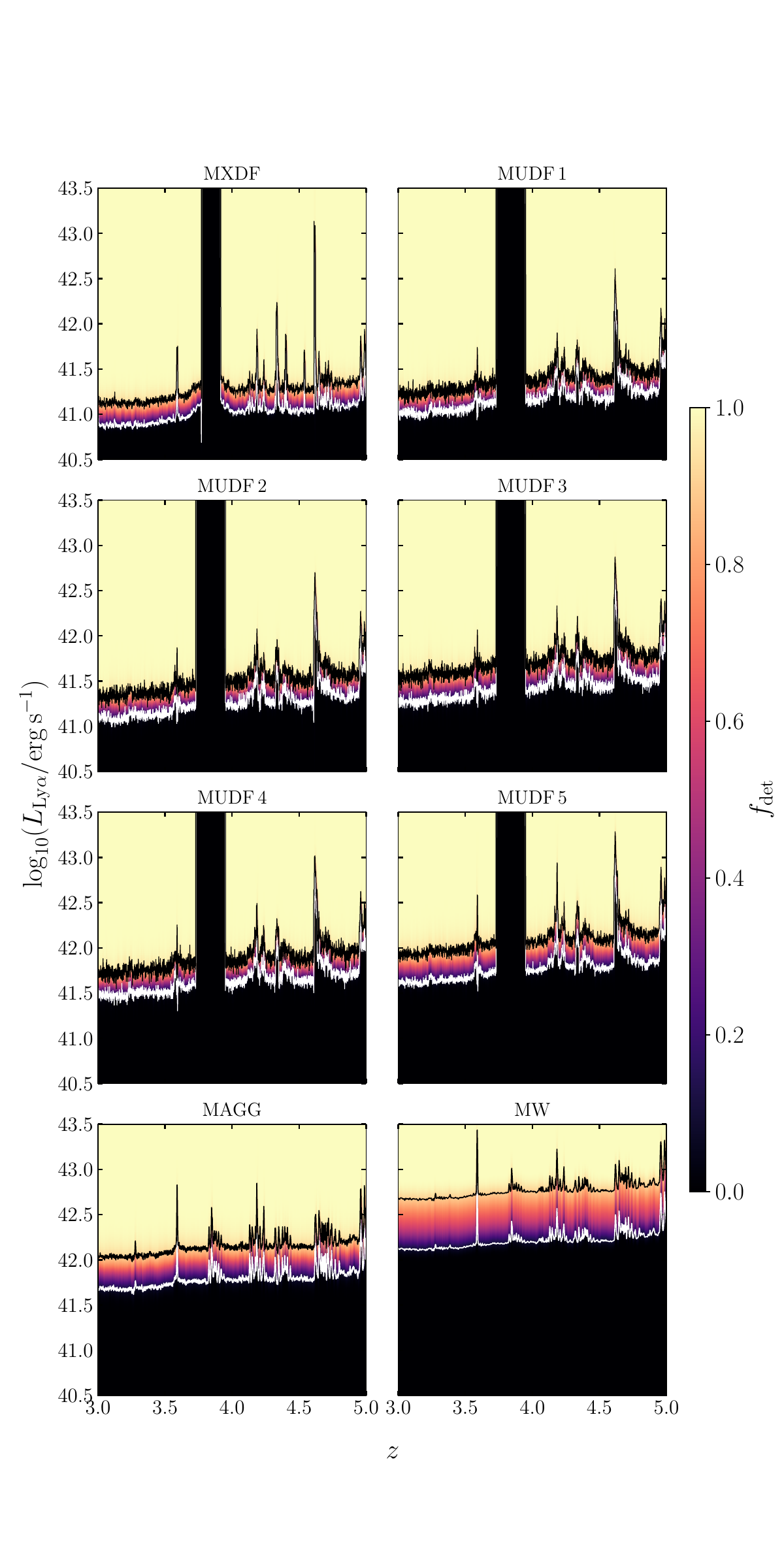}
\caption{Selection functions $P_\mathrm{det,k}(\tilde{L}, z)$ for LAEs analysed in each individual survey. The white and black contours indicate the $10\%$ and $90\%$ completeness limits, respectively. The MW and MAGG selection functions are taken from \cite{Herenz2019} and \cite{Fossati2021}, respectively. The vertical stripe with a detection fraction of zero around redshift $z \sim 3.8$ in the MXDF and MUDF surveys is caused by the Ground Layer Adaptive Optics module, which uses an artificial laser guide star to improve image quality during the observations \citep{Fossati2019, Bacon2023}.
This region therefore contains no data.}
\label{fig:sel_funcs}
\end{figure}

\end{appendix}

\end{document}